# Quantum interactive learning tutorial on the double-slit experiment to improve student understanding of quantum mechanics

Ryan Sayer,[1] Alexandru Maries,[2] and Chandralekha Singh[3]

[1]*Department of Physics, Bemidji State University, Bemidji, Minnesota 56601, USA*
[2]*Department of Physics, University of Cincinnati, Cincinnati, Ohio 45221, USA*
[3]*Department of Physics and Astronomy, University of Pittsburgh,
Pittsburgh, Pennsylvania 15260, USA*

Learning quantum mechanics is challenging, even for upper-level undergraduate and graduate students. Research-validated interactive tutorials that build on students' prior knowledge can be useful tools to enhance student learning. We have been investigating student difficulties with quantum mechanics pertaining to the double-slit experiment in various situations that appear to be counterintuitive and contradict classical notions of particles and waves. For example, if we send single electrons through the slits, they may behave as a "wave" in part of the experiment and as a "particle" in another part of the *same* experiment. Here we discuss the development and evaluation of a research-validated Quantum Interactive Learning Tutorial (QuILT) which makes use of an interactive simulation to improve student understanding of the double-slit experiment and strives to help students develop a good grasp of foundational issues in quantum mechanics. We discuss common student difficulties identified during the development and evaluation of the QuILT and analyze the data from the pretest and post test administered to the upper-level undergraduate and first-year physics graduate students before and after they worked on the QuILT to assess its effectiveness. These data suggest that on average, the QuILT was effective in helping students develop a more robust understanding of foundational concepts in quantum mechanics that defy classical intuition using the context of the double-slit experiment. Moreover, upper-level undergraduates outperformed physics graduate students on the post test. One possible reason for this difference in performance may be the level of student engagement with the QuILT due to the grade incentive. In the undergraduate course, the post test was graded for correctness while in the graduate course, it was only graded for completeness.

## I. INTRODUCTION

According to a poll of Physics World readers, the interference of single electrons in a double-slit experiment is "the most beautiful experiment in physics" [1]. The beauty of this experiment comes from its powerful illustration of the quantum nature of microscopic particles. This experiment (schematic diagram of the experimental setup shown in Fig. 1) is useful for helping students learn about foundations of quantum mechanics, including the wave-particle duality of a single particle, the probabilistic nature of quantum measurements, collapse of the wave function upon measurement, etc. It illustrates how information about which slit a particle went through, or "which-path information" (WPI), can destroy the interference pattern on the distant screen when a large number of single particles are sent [2,3]. Prior research on student learning of quantum mechanics has found that many students struggle with foundational concepts in quantum mechanics after traditional instruction and many tools have been developed which can help improve student understanding of these concepts [4–47]. However, there are few experiments that elucidate the basic principles of quantum mechanics as clearly as the double-slit experiment. Here, we discuss the development and evaluation of a research-validated interactive tutorial designed to help students develop a good grasp of foundational issues in quantum mechanics in the context of the double-slit experiment (DSE).

The development and use of research validated tools to help students learn upper-level quantum physics has been a subject of continuing interest. Our group has investigated difficulties students have in learning various concepts in upper-level quantum mechanics, and developed and evaluated research-validated interactive tutorials or Quantum Interactive Learning Tutorials (QuILTs) [49]. The use of research-validated QuILTs in upper-level quantum mechanics courses shows that they help students develop a good grasp of quantum mechanics concepts [50,51]. The QuILTs use a guided approach to learning and often incorporate interactive simulations. They are structured in a way which allows students to make predictions and observe the outcome of an experiment in a computer simulation, after which they are guided to reconcile the difference between what they predict and what they observe and extend and repair their knowledge structure. In other words, students are asked to compare their observations with their predictions, and if their predictions do not agree with the simulation, they are given scaffolding support and feedback to reconcile the differences. The QuILTs provide students with appropriate guidance and prompt feedback as they strive to extend, organize, and repair their knowledge structure related to foundational issues in quantum mechanics using concrete examples. Previous QuILTs have been developed on topics such as the possible wave function, bound state and scattering state wave functions, time development of wave functions, the uncertainty principle, the Stern-Gerlach experiment, quantum key distribution, quantum measurement, Larmor precession of spin, addition of angular momentum, and the Mach-Zehnder interferometer with single photons and quantum eraser [50–56].

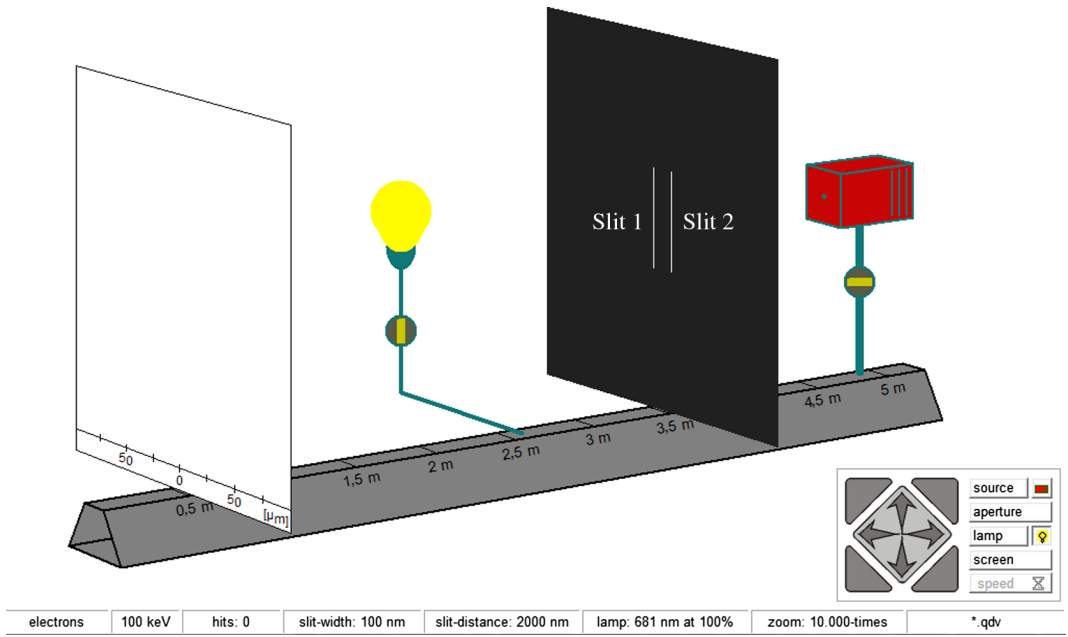

FIG. 1. The basic setup of the double-slit experiment with single particles, consisting of a particle source, a plate with two narrow slits (labeled slit 1 and slit 2), a monochromatic lamp (light bulb) placed near the two slits, and a screen that detects the particles. Figure adapted from simulation developed by Klaus Muthsam [48].

Here, we discuss the development and evaluation of a research-validated QuILT on the DSE involving single particles sent one at a time through the slits [2]. We first provide a background on the relevant concepts in the DSE, after which we discuss theoretical frameworks which inform our investigation. Next, we discuss common student difficulties that were identified related to the DSE with single particles sent one at a time through the slits, describe how the DSE QuILT was developed using research as a guide, and summarize its structure. We then discuss the evaluation of the QuILT including its effectiveness in addressing common student difficulties.

## II. BACKGROUND

Before we discuss research on student conceptual difficulties with the DSE and how that research informed the development and evaluation of a QuILT on the DSE, we provide a brief background on the DSE relevant for the QuILT (setup shown in Fig. 1). In particular, we discuss how one may reason in terms of which-path information [2,3] to predict the pattern observed on the screen after a large number of single particles are emitted from the source. In this setup, the particle source emits single particles one at a time towards a plate with two narrow slits that are finally detected on the distant screen. We will use electrons for this discussion, but the reasoning discussed can be applied to any other particle that is sufficiently small (e.g., protons, neutrons, Na atoms, etc.) to create an interference pattern under appropriate conditions with this setup. A monochromatic lamp is placed between the slits and the screen that emits photons that can scatter off the electrons, and, for simplicity, it is assumed that if scattering does occur, it occurs at or very near the slits, and that an electron only scatters a single photon, i.e., multiple scattering is neglected. This thought experiment is described in detail in Feynman's lectures [2]. We assume that the parameters of the experiment, e.g., the distance between the narrow parallel slits and wavelength of the electrons are such that when the monochromatic lamp is turned off, an interference pattern is observed on the screen after a large number of electrons are detected. When the lamp is turned on, it emits photons of a certain wavelength that scatter off the electrons and the intensity of the lamp can vary from 0% to 100%, where 100% means that all of the electrons at the slits scatter off photons emitted by the lamp. Scattering between a photon and an electron corresponds to a measurement and it can localize the electron's position depending upon the wavelength of the photon emitted by the lamp. In other words, the scattering process localizes the electron in a region of length scale comparable to the wavelength of the photon. Therefore, if the wavelength of the photon is significantly smaller than the distance between the slits, since we are assuming that the scattering process occurs at the slits, the scattering will provide information about the position of the electron during the scattering process: at one slit or at the other slit. This is what is referred to as "having WPI": knowing that the electron went through one slit or the other, but not both. In this case, if the intensity of the lamp is 100%, the interference pattern that would otherwise be observed on the screen (when the lamp is turned off) is destroyed due to the scattering between an electron and a photon emitted by the lamp when the lamp is turned on. If the lamp is of intermediate intensity, say 50%, only half of the electrons scatter off of photons and do not interfere, whereas the other half do interfere. Therefore, the pattern observed on the screen after a large number of electrons are detected will be an interference pattern (50% of electrons that do not scatter) on top of a uniform background due to the 50% of the electrons that do not interfere (so overall, there will be a reduced contrast in the interference pattern).

If instead the wavelength of the photons is significantly larger than the distance between the slits, scattering between an electron and a photon does not provide WPI. In this case, the electron is localized in a region of length scale significantly larger than the distance between the slits, and therefore it is not possible to know that the electron goes through one slit or the other. In this case, the electron goes through both slits and we observe an interference pattern on the screen indistinguishable from when the lamp is turned off. Furthermore, the intensity of the lamp is irrelevant because regardless of whether an electron scatters off a photon or not, it will still interfere with itself.

For a more precise definition of WPI in the context of DSE, see[1] footnote 1.

## III. THEORETICAL FRAMEWORKS INFORMING THE INVESTIGATION OF STUDENT DIFFICULTIES AND DEVELOPMENT OF THE QUILT

Research on student reasoning difficulties in learning upper-level quantum mechanics is inspired by cognitive theories that highlight the importance of knowing student difficulties in order to help them develop a functional understanding. According to the cognitive apprenticeship model, students can learn relevant concepts and develop effective problem-solving strategies if the instructional design involves three essential components: modeling, coaching and scaffolding, and weaning [57]. In this approach, "modeling" means that the instructor uses tools to demonstrate and exemplify the skills that students should learn (e.g., how to solve physics problems systematically). "Coaching and scaffolding" means that students receive appropriate guidance and support as they actively engage in learning the skills necessary for good performance. "Weaning" means gradually reducing the support and feedback to help students develop self-reliance.

In traditional physics instruction, especially at the college level, there is often a lack of coaching and scaffolding [58,59]. Instructors typically give a lecture explaining the topics and demonstrate how to solve a few example problems. Students are then told to practice applying the skills on their own on homework with no guidance and little feedback (except for correct or incorrect after turning in the homework). Additionally, years of teaching experience and practice often make the instructor's reasoning and problem-solving skills implicit: they no longer have to think about what they are doing at each step, which is a hallmark of expertise. This suggests that as they lecture to students, they may be deficient in modeling effective problem solving because they are no longer explicitly aware of their problem-solving skills which have become automatic. In other words, students are often expected to learn and apply the expertlike practices not modeled explicitly by their instructors when working on the homework problems on their own. This situation is akin to a piano instructor demonstrating for the students how to play a particular musical piece and then asking students to practice on their own. The lack of prompt feedback and scaffolding support can be detrimental to learning. Advanced students are still developing expertise in quantum mechanics, and they need coaching and prompt feedback in order to develop expertise and build a robust knowledge structure. Research-validated QuILTs, which use a guided inquiry-based approach to learning, can provide students the opportunity to receive coaching and scaffolding as they engage in a guided exploration of quantum physics concepts.

---

[1]The concept of WPI at a detector (such as a screen) is useful when the state of the system is in a superposition of two different spatial path states as in the DSE. In general, when a detector can project both components of the path state, then WPI is unknown. On the other hand, when a detector can only project one component of the path state, then we have complete WPI. For example, a single electron that is delocalized in space can go through both slits before reaching the screen and interfere with itself. In this case, we do not have WPI for the electron, and interference of single electrons is observed on the screen. In other words, interference occurs because, as the electron wave function evolves when the electron travels from the slits to the screen, the two components related to the different path states pick up different phases that are related to the path lengths—distances from one slit (or the other slit) to the point on the screen where it is detected. Depending on the path length difference, the probability of detecting the electron (corresponding to the absolute value of the wave function squared) depends on $\cos(\Delta\phi)$ (where $\Delta\phi$ is the phase difference) which results in an interference pattern. However, if we measure which slit the electron went through, the wave function collapses to one or the other path state at the slits, and when the electron reaches the screen, the detector (screen) can only project that particular path state. Therefore, the absolute value of the wave function squared does not depend on the path length difference between the two components of the wave function, and no interference is observed. In this case, by measuring the electron near one of the slits, we obtain WPI for the electron and it therefore cannot interfere with itself when it reaches the screen.

Moreover, Schwartz, Bransford, and Sears' framework of "preparation for future learning" (PFL) suggests that effective instructional design that focuses on developing adaptive expertise should include elements of both innovation and efficiency [60]. While there are many interpretations of the PFL framework, efficiency and innovation can be considered two orthogonal dimensions in the instructional design. If the instructor focuses only on helping students develop efficiency through repeated practice involving a high measure of consistency, cognitive engagement will be diminished and learning will be less effective (students are likely to become routine experts) [60]. Conversely, if the instructional design is solely focused on innovation, students will struggle to connect what they are learning with their prior knowledge and learning and transfer will be inhibited (students are likely to become frustrated novices) [60]. One example of instruction high on the efficiency dimension would be an instructor solving an example problem in class and giving students similar problems with numbers changed to do on their own. Students can easily apply the same procedure correctly to a similar situation without developing their ability to transfer their learning to other situations. In contrast, an example of instruction high on the innovation dimension would be something like discovery learning. For example, students may be given batteries, wires, and light bulbs and asked to figure out the rules governing current and voltage without any more guidance when they do not have the prior knowledge in this area to make progress. In this case, the lack of guidance along with the cognitive difficulty of the task can make the task very frustrating for students who will have difficulty learning from the task.

Based upon this framework, in order to help students become adaptive experts, incorporating the elements of efficiency and innovation into an instructional design demands that the instruction challenges students to work through and learn from tasks somewhat above their current understanding by building on their existing knowledge and level of expertise. Innovation and efficiency are both incorporated in the guided inquiry-based active-learning approach in the QuILT. The efficiency aspect is incorporated in the QuILT in that instead of being asked to engage in discovery learning, the guided inquiry-based learning sequences in the QuILT are designed based upon a careful cognitive task analysis of the underlying knowledge that students should learn and the types of common difficulties students have in learning these concepts. The innovation aspect is incorporated by ensuring that students are challenged to think through carefully designed learning sequences, e.g., which require them to reason using WPI in novel situations. The QuILT strives to provide sufficient coaching and scaffolding to allow students to make progress and build a good knowledge structure while keeping them actively engaged in the learning process.

## IV. METHODOLOGY FOR INVESTIGATION OF STUDENTS' CONCEPTUAL DIFFICULTIES

The QuILT was used in an upper-level undergraduate quantum mechanics course (and a graduate-level TA training course) because at the institution where the study was carried out, DSE is discussed in the modern physics course only for the basic setup. The main focus of the QuILT is on helping students understand more complex setups which include a monochromatic lamp placed between the slits and the screen. Our discussion with a faculty member who regularly teaches the modern physics course indicated that the QuILT was not appropriate for his class. Also, a look at modern physics textbooks (e.g., Serway, Moses, and Moyer [61]) confirms instructors' views and there are no discussions of DSE setups with single particles which include a monochromatic lamp placed between the slits and the screen. Therefore, the QuILT was utilized in an upper-level quantum mechanics course (or courses at higher levels). These students were all familiar with the basic setup of DSE before the more complex setups were discussed.

To identify student conceptual difficulties involving the DSE, we surveyed the literature, in particular, for research on student understanding of the DSE as described in both typical modern physics and quantum mechanics courses. Based on the difficulties identified in the literature and on the concepts involved in the DSE setup shown in Fig. 1, we drafted open-ended questions and administered them to upper-level undergraduate and graduate students in various quantum mechanics classes after instruction in relevant concepts. Based upon student responses to questions in one quantum mechanics course, some of the written open-ended questions were revised to further probe student understanding. In addition, individual interviews were conducted with students in these courses or with graduate students after traditional instruction in relevant concepts. The initial questioning and the data collected helped us construct a pretest and post test to quantify these difficulties. This research on student difficulties was also used as a guide for developing and refining the QuILT, which uses a guided inquiry-based approach. Since the purpose of the original open-ended questions and interviews was to obtain a list of common student difficulties and qualitative reasoning, in the next section we only describe student difficulties qualitatively and discuss quantitative data later when we discuss pretest and post test results.

In the undergraduate course in which the study was carried out, traditional instruction included topics such as the de Broglie relation, calculation of the de Broglie wavelengths of different particles, an overview of the patterns that form on the distant screen in the DSE after a large number of single particles are sent one at a time through the slits, and a brief overview of the relevance of the information about which slit the particle went through to whether an interference pattern is observed on the screen. The questions on the pretest and post test were graded using rubrics which were designed to assess student understanding of relevant concepts by considering responses for multiple questions together (an example of a specific question is provided later). A subset of the responses for all questions (20%–30%) was graded separately by two investigators. After comparing the grading of some students, the raters discussed any disagreements in grading and resolved them so that the interrater agreement after the discussions was better than 90%.

We conducted approximately 85 h of individual interviews before, during, and after the development of different versions of the DSE QuILT and the corresponding pretest and post test. The interviews used a semistructured, think-aloud protocol [62] and were designed to provide the researchers with a better understanding of the rationale students used to answer foundational questions related to the DSE. During the semistructured interviews, upper-level undergraduate and graduate students were asked to verbalize their thought processes while answering the questions. Students read the questions related to the DSE setup and answered them to the best of their ability without being disturbed. They were prompted to think aloud if they became quiet for a long time. After students had finished answering a particular question to the best of their ability, they were often asked to further clarify and elaborate issues that they had not clearly addressed earlier.

Below, we first provide a review of the difficulties identified via surveying the literature and then discuss the difficulties identified in our research with physics undergraduate and graduate students who had traditional instruction in upper-level quantum mechanics.

## V. STUDENT DIFFICULTIES

Student difficulties with several concepts relevant for understanding the DSE have been investigated by several researchers, for example, the nature of the electron [63–65], particle-wave duality in the context of diffraction experiments [65], and the DSE [66–69], while others have investigated student understanding of the wave nature of light [70,71]. These investigations utilized diverse groups of students from high school to advanced undergraduates (e.g., in upper-level quantum mechanics course). With regards to the nature of the electron, the majority of preuniversity students (who have had instruction in the particle-wave duality of light) visualized the electron as some sort of particle, e.g., a small ball or sphere, thus favoring the Bohr model of the atom [63–65]. This finding has implications for students' wave-particle duality conceptions. Moreover, when presented with the results of a diffraction experiment for electrons, many students (upwards of 2/3) who previously have studied such effects for light, recognized that the electron also exhibits wave characteristics, but they still held ideas related to the particle nature of the electron (e.g., it moves in straight trajectories) and sometimes attempted to reconcile wave and particle behavior by commenting that the electron (as a localized particle) moves along the trajectory of a wave (i.e., sinusoidally) [65].

For DSE, when electrons are sent one at a time, Thacker found [66] that the majority of high-school students she interviewed thought that the pattern on the screen would consist of two dots, i.e., that electrons would exhibit a particle behavior. College students in a modern physics course would often explain the DSE pattern observed by electrons bouncing into one another (particle nature). These students were aware of the particle-wave duality of light and x rays and even commented that "maybe electrons had wavelike properties", but they could not correctly describe the wavelike nature of an electron [66]. These examples illustrate that students often try to apply classical concepts they are familiar with to the unfamiliar context of quantum mechanics phenomena.

Akiskaien and Hirvonen have found similar results [67] when investigating in- and preservice teachers' conceptions of the DSE. Nearly all the teachers in the study had taken a quantum and atom physics course that discussed wave-particle duality among other quantum topics. Although the majority of the teachers had the correct interpretation of the DSE with electrons, the most common incorrect answers described patterns consistent with the particle nature of electrons. They also found that in-service teachers with experience teaching modern physics were more likely to accurately describe the interference pattern observed in a DSE with electrons.

In a study with students taking a modern physics course, Ambrose et al. [68] found common incorrect ideas coming from attempting to apply classical physics ideas to quantum mechanics, for example, believing that electrons travel in straight or sinusoidal paths, or that their trajectories bend at the slit edges. Many students also believed that more than one electron is needed for interference, i.e., one electron needs at least one other electron to interfere with. As a result of identifying these difficulties and others related to recognizing the effect of changes of certain parameters (e.g., slit width, energy of incoming electrons, etc.) in experiments like the DSE [69], Vokos et al. developed a tutorial on the wave properties of matter which was found to be effective in helping students learn these challenging concepts.

In another study, Mannila et al. [72] presented students enrolled in an advanced undergraduate quantum mechanics course with pictures showing the gradual buildup of a double-slit electron interference pattern. After being shown the statistical formation of an interference pattern in the case of electrons, students were asked to explain how this pattern arises. The majority of students gave answers that indicated that they had trouble reconciling the new quantum ideas they had learned with their classical intuition evidenced. For example, many students mentioned that electrons are localized particles that follow certain trajectories and their wavelike characteristics makes them interfere. As another example, some students believed that "the wave function is […] a guiding wave or 'ghost-wave' governing the trajectory of the particle" [72]. Less than 30% of the students discussed the statistical interpretation of the experiment, although even some of these students focused on ideas related to trajectories followed by single electrons.

However, we are unaware of prior studies reporting on common student difficulties with the DSE with single particles that includes a monochromatic lamp placed near the slits as in the present investigation. Therefore, we now turn to discussing common student difficulties identified in the present investigation which will be elaborated upon later while discussing the pretest and post test results.

### A. Difficulty reasoning in terms of which-path information

Before working on the DSE QuILT, it is unlikely that students have any knowledge of the concept of WPI, and, in our interviews, none of the students tried to use WPI reasoning to answer questions about interference in the DSE setup with the monochromatic lamp. There is no analogue to the concept of WPI in classical mechanics, and many students find it difficult to reconcile their classical intuition with the quantum effects observed in the DSE. The concept of WPI and its relation to the interference at the screen in the DSE can be difficult for students if they are not given appropriate scaffolding support as they learn these counterintuitive concepts. Furthermore, even when discussing the basic DSE setup (without the lamp), some students explained single electron interference by stating that one electron going through one slit interferes with another electron going through the other slit, even though they have been told at the beginning that a single electron is sent at a time. This type of reasoning suggests that students have difficulty understanding the wave nature of electrons; similar difficulties have been found in prior investigations [66,67] with students in modern physics courses. In other words, this concept of single electron interference is so difficult for students to grasp that they ignore relevant information provided (one electron at a time) and explain it in their own way consistent with their mental model.

### B. Difficulty recognizing the effect of the wavelength of the photons emitted by the lamp on the interference pattern

Many students struggle to incorporate the wavelength of the lamp's photons into their responses to the DSE questions before working on the QuILT. In interviews, students were asked to predict the pattern that will be observed on the screen in a DSE when the wavelength of the photons is significantly smaller than the distance between the slits. Many students claimed that the wavelength of the scattered photons is not important, and that only the intensity of the lamp matters. For example, in an interview, when asked explicitly about why he did not incorporate the wavelength of the scattered photon, one student simply noted that he thought that the answer to what happens to the interference pattern should be independent of the photon wavelength and only depend on how many photons are interacting with the single particles incident on the slits (the student thought that every incident particle that interacts with a photon will not show interference regardless of the photon's wavelength).

Some students also struggled to differentiate the wavelength associated with particles such as electrons or atoms emitted by the particle source from the wavelength associated with the photons emitted by the lamp. This lack of differentiation led some students to claim that if a photon emitted by the lamp and a particle emitted by the particle source have the same wavelength, they can interfere with each other destructively and annihilate each other. For example, when asked to describe a situation in which the presence of the lamp will lead to the destruction of the interference pattern on the screen, some students described a scenario in which the photon and the particle from the slit destructively interfere with each other. Discussions suggest that these students were familiar with the concept of a particle behaving as a wave but had not yet developed a deeper understanding to realize that an electron and a photon with the same wavelength but with opposite phase cannot destructively interfere with each other. When asked in an interview to explain his reasoning, one student simply asserted that he feels this way because "that's just what I've been told [by his instructor in class]". This type of response suggests that even advanced students are likely to misinterpret what they learn from lectures particularly if what the instructor tells them is not consistent with their existing knowledge structure. Moreover, this type of response also conveys an epistemology about learning quantum physics in which the advanced student views the instructor as an authority figure and accepts what the instructor says without questioning or making sense of it and integrating it with his existing knowledge structure.

### C. Difficulty recognizing that lamp intensity alters the pattern observed on the screen only when the wavelength of the photons is significantly smaller than the distance between the slits

This difficulty is a corollary of the preceding difficulty with the impact of the wavelength of the photon on the interference pattern on the screen. As discussed in Sec. II, when the wavelength of the photons emitted by the lamp is significantly smaller than the distance between the slits, the intensity of the lamp will determine the fraction of incoming particles for which WPI is available (e.g., 50% intensity means that 50% of the incoming particles will not interfere and 50% will interfere). If instead the wavelength of the photons is significantly larger than the distance between the slits WPI is not available for any of the particles and full interference is observed. Since the effects of photon wavelength and lamp intensity are intertwined, it is challenging to disentangle students' difficulties as pertaining to one thing or the other. Interviews suggest that due to the preceding difficulty with the wavelength, many students expect that changing the intensity will affect the interference pattern regardless of the wavelength of the photons emitted by the lamp. Students were asked in individual interviews to predict the pattern that will form on the screen in the case in which the wavelength of the photons was much larger than the slit separation and the intensity of the lamp was initially 100% and then reduced to 50%. In response to this question, many of the students predicted that the patterns observed on the screen would be different in the two cases. For example, when one student was asked to explain why he predicted different patterns in the two cases, instead of explaining a causal relation of some kind, he emphatically stated "It [the pattern] HAS to change in some way" (emphasis on "has"). This type of a response from advanced students in the context of quantum mechanics illustrates a powerful phenomenological primitive that many students have [73], that when you change the input of a system, the output must always change in some way in response. The DSE QuILT therefore uses a guided inquiry-based approach and strives to provide scaffolding support in order to first help students recognize that the wavelength of the photon is an important consideration in determining whether interference is observed and then using that fact to reason about how the intensity of the lamp influences the interference pattern (because their gut instinct is to expect that a change in the DSE setup, i.e., varying the intensity of the lamp must result in a change in the interference pattern). The guidance was designed to help students recognize how changing the wavelength of the photons affects the interference pattern before any additional complexities are added with varying the lamp intensity.

## VI. DEVELOPMENT OF THE QUILT, ITS STRUCTURE AND LEARNING OBJECTIVES

### A. Development and validation of the DSE QuILT

The difficulties discussed above indicate that upper-level undergraduate and graduate students may either have difficulties with the concept of resolution or they may struggle to develop a coherent understanding of the foundational issues in quantum mechanics relevant for understanding whether interference will be observed in the DSE under various conditions. These students can benefit from a research-validated tutorial which uses a guided inquiry-based approach to help them learn these concepts involving single particles passing through a DSE. Therefore, we were motivated to develop a research-validated QuILT on the DSE with single particles.

The development of the QuILT was a cyclical, iterative process which included the following stages: (i) development of a preliminary version of the QuILT based upon a cognitive task analysis of the underlying concepts and knowledge of common student difficulties found via research; (ii) implementation and evaluation of the QuILT by administering it to individual students, asking them to think aloud as they worked on it, and measuring improvement via their performance on pretests and post tests; and (iii) after determination of its impact on student learning and assessment of what difficulties were not adequately addressed by a particular version of the QuILT, making refinements and modifications based upon the feedback from the implementation and evaluation of the previous version.

Different versions of the QuILT were also iterated several times with five physics faculty members to ensure that experts agreed with the content and wording. The faculty feedback complemented the feedback obtained by having advanced students work on the QuILT in individual think-aloud interviews. These interviews helped to ensure that the guided approach was effective and the questions were unambiguously interpreted by students, as well as to better understand students' reasoning as they answered the questions. A total of approximately 85 h of individual interviews were conducted with students during the development and assessment phases of the DSE QuILT.

### B. Structure of the DSE QuILT

The guided inquiry-based approach used in the DSE QuILT strives to help students build on their prior knowledge and accounts for common student difficulties to help them develop a good knowledge structure of foundational issues in quantum mechanics using the context of the DSE. The QuILT consists of these components to be used in the following order: a pretest, a warm up, a main tutorial, an associated homework component, and a post test, as shown in Fig. 2. The pretest consists of free-response questions involving the DSE with single particles and a monochromatic photon source placed between the slits and the screen. The photon source emits photons of a particular wavelength that scatter off the single particles at the slits. The warm up serves to help students learn about the double-slit experiment without the photon source placed between the slits and the screen and focuses on the de Broglie relation, wave-particle duality as manifested in the DSE, how the registering of a particle on the distance screen can be viewed as a measurement of position, and the impact of measurement on the wave function of the particle. The warm up helps prepare students to learn the prerequisite concepts required to engage effectively with the main tutorial in the sequence. Students work on the QuILT in class in groups, and whatever they do not finish in class, they work on at home. After working on the main tutorial, which is conceptual in nature, students work on a homework component that connects the conceptual and mathematical aspects of the DSE to help students connect the conceptual and quantitative aspects of quantum mechanics involved in the experiment [2]. Finally, students work on a post test that is identical to the pretest. We note that giving the same post test as the pretest to evaluate the extent to which students have learned a certain topic is common practice in physics education research; prior research with the Force Concept Inventory, for example, has found that giving the pretest does not affect post test results [74]. Additionally, the pretest and post test used here include only free-response questions and it is very challenging to make different questions that are well matched to the learning objectives (described in the question below). Moreover, the pretest was never returned to the students.

The warm up and main tutorial make use of a computer simulation in which students can manipulate the DSE setup and observe the resulting pattern on the screen. This setup involves a plate with two slits, a particle source, a screen which serves as the detector for the experiment, and a photon source (light bulb) placed near the two slits, as shown in Fig. 1. Students are asked to predict the pattern that will appear on the screen based on the type of particles emitted by the source, their energy, the width and separation of the two slits, the wavelength of the photons emitted by the photon source, and the intensity of the photon source. Students then use the simulation to check their predictions. When a large number of particles has reached the screen, students can observe an interference pattern consisting of several dark and bright fringes or a featureless distribution without any interference fringes. Figure 3 shows a screenshot of the simulation in which an interference pattern has formed on the screen. Students can use the computer simulation to verify that there are interference fringes on the screen when the chosen parameters are used (as shown in Fig. 3). Figure 4 shows a screenshot of the simulation in which no interference pattern has formed on the screen after a large number of particles has reached it. Students can also observe a combination of the two in which the dark and bright fringes are still visible but they are on top of a uniform background of scattered particles that arrive at the screen (in which case, there is WPI for some photons but not for others and there is a reduced contrast in the interference pattern due to some photons, for which WPI is known, not displaying interference). Students are then given an opportunity to reconcile the difference between their predictions and observations before proceeding further in the tutorial. They are also provided checkpoints to reflect upon what they have learned and to make explicit connections with their prior knowledge.

### C. DSE QuILT learning objectives

The DSE QuILT focuses on helping students learn about interference of single particles in a DSE with a photon source placed near the two slits. In particular, the DSE QuILT strives to address common student difficulties and help students develop a good knowledge structure of the foundational quantum mechanical concepts involved (such as wave-particle duality, quantum measurement and collapse of wave function) in a concrete context by focusing on the following learning objectives:

**Learning objective 1.**—Predict the number density after $N$ particles have been detected at the screen in cases in which WPI is known or unknown and make connections between WPI, the number density, and the absence or presence of interference.

As discussed in Sec. III, many students struggled with reasoning in terms of WPI, which is a convenient conceptual framework for considering whether interference is observed in a particular situation or not. The QuILT first provides guidance to help students learn to predict the number density for the case in which WPI is known for single particles sent through the slits. The students are then provided scaffolding support and appropriate feedback for the case in which WPI is not available and determine the number density on the screen for this case. This approach helps students make connections between the number density and whether WPI is known or not. They are then asked to identify the number density for which interference will be observed and the number density for which interference will not be observed on the screen. Thus, the guided approach strives to help students make connections between (i) whether WPI is known or not, (ii) the number density after a large number of particles reach the screen, and (iii) the presence or absence of interference fringes.

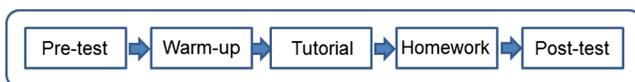

FIG. 2. Sequence of components comprising the entire DSE QuILT suite.

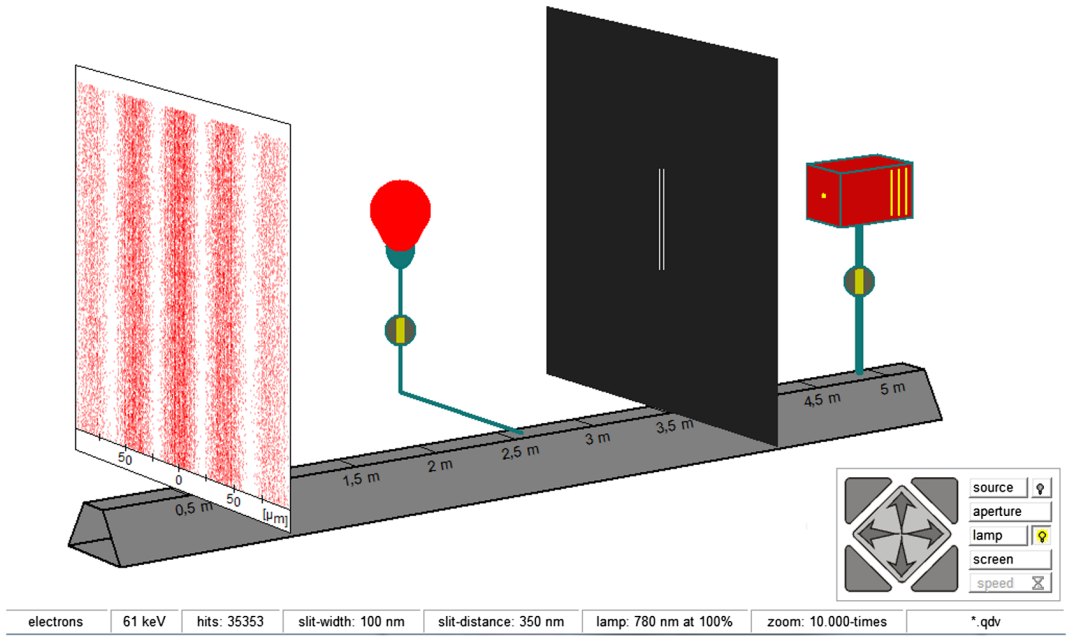

FIG. 3. Screenshot of the computer simulation of the DSE (which is part of the QuILT) for a situation in which an interference pattern has formed on the screen after a large number of single particles have been sent through the slits to the screen. In the simulation, the slit width is 100 nm, the distance between the slits is 350 nm and the kinetic energy of the electrons is 61 keV. Figure adapted from simulation developed by Klaus Muthsam [48].

For example, in order to scaffold student learning, the following question in the QuILT asks students to think about what changes occur in the number density of particles on the screen based on whether WPI is available for the particles incident on the slits. [Note that in the notation used in the QuILT, $\Psi_1(x)$ and $\Psi_2(x)$ represent the wave function at point $x$ on the screen when slit 2 or slit 1 is closed, respectively, and $\Delta\phi$ represents the phase difference between $\Psi_1(x)$ and $\Psi_2(x)$ at point $x$ on the screen. When both slits are open, students must take into account both $\Psi_1(x)$ and $\Psi_2(x)$.]

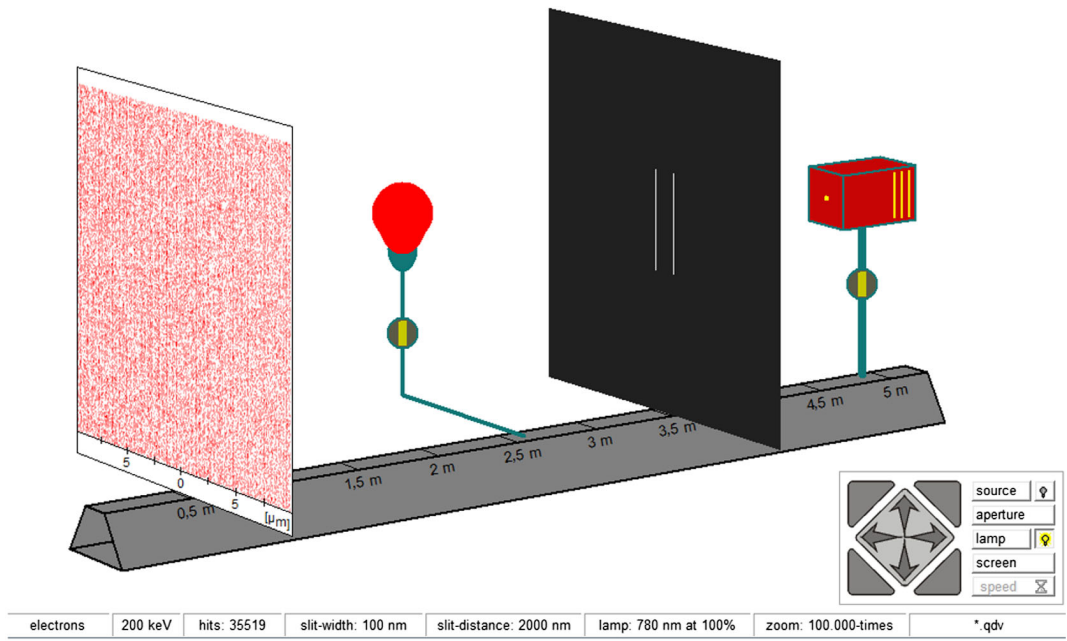

FIG. 4. Screenshot of the computer simulation of the DSE for a situation in which no interference pattern is formed on the screen after a large number of single particles have been sent through the slits to the screen. In the simulation, the slit width is 100 nm, the distance between the slits is 2000 nm and the kinetic energy of the electrons is 200 keV. Figure adapted from simulation developed by Klaus Muthsam [48].

Circle all of the following statements about the double-slit experiment that are correct:

  (i) If the cross term $[2|\Psi_1(x)||\Psi_2(x)|\cos\Delta\phi]$ in the expression for the expected number density of electrons is negligible in a given situation, interference effects will be negligible.
  (ii) If we obtain which-path information, i.e., information about which slit the electron went through, the cross term in the expression for the expected number density of electrons vanishes.
  (ii) If we first square the wave function from each slit and then add the results to obtain the total probability density for a single electron, i.e., $|\Psi(x)|^2 = |\Psi_1(x)|^2 + |\Psi_2(x)|^2$, and then sum over all electrons to obtain the expected number density of electrons at each point $x$ on the screen, we would conclude that there are no interference effects.

Explain your reasoning.

We note that thus far, students are only considering the basic DSE setup in relatively straightforward setups in which WPI is known (when one slit or the other is blocked) or unknown (neither slit is blocked). These simpler setups help them make the appropriate connections between WPI, number density, and interference. Learning objective 2 focuses on helping students recognize other situations in which WPI is known (e.g., monochromatic lamp which emits photons with wavelength significantly smaller than the distance between the slits) and therefore the QuILT gradually builds in complexity while guiding students to make appropriate connections.

*Learning objective 2.*—Predict the qualitative features of the pattern that will form on the screen after a large number of particles have been sent through the slits depending on the wavelength of the photons that scatter off the particles.

After the QuILT guides students to reason about WPI and how to incorporate it into the DSE to predict the pattern that forms on the screen after a large number of single particles are detected at the screen (without a monochromatic lamp between the slits and the screen), students learn about the role of photon wavelength in determining WPI for the particles incident on the slits. Then, the QuILT builds upon students' understanding of WPI by incorporating the simulation. The QuILT asks students to make predictions about the pattern that will form on the screen after a large number of particles reach the screen based upon the wavelength of the photons from the lamp.

Students are then asked to use the simulation to check their predictions. If the simulation does not agree with their predictions, students must reconcile the difference by reconsidering their reasoning when making the prediction. When students work in small groups in class, they discuss their predictions and observations with their peers. The QuILT then provides guidance and support to help them develop a good understanding of these issues. They are guided to use different simulations, some of which have photon wavelengths significantly smaller than the distance between the slits and some of which have photon wavelengths significantly larger than the distance between the slits. Many students do not realize that the wavelength of the photons is related to the length scale over which the scattered photons can resolve an object. In the tutorial, they consider different DSE setups as described above, use the simulation, answer questions regarding the issue of how photon wavelength affects the interference pattern, and, finally, go over a checkpoint which summarizes the main concepts involved. This approach strives to help students grasp how photon wavelength affects the interference pattern. We note that although we often discuss using single electrons in the DSE here, students learn that everything that follows can be applied in a very similar manner to other particles. The QuILT included other particles, and the source used in the simulation could be used to select between electrons, Na atoms, muons, etc. The following question in the QuILT uses a hypothetical conversation between three students to scaffold student learning about the role of photon wavelength in the DSE:

Consider the following conversation between Pria, Mira, and Nancy about why an important consideration in the loss of the interference fringes is the comparison of the slit separation with the wavelength of the photons emitted by the lamp.

- Pria: I think that we will always have which-path information regardless of the wavelength of the photons emitted by the lamp as long as the lamp has high intensity. If the lamp has high intensity, virtually every electron will scatter off one photon. Therefore, we will be able to determine where each electron scattered from (which slit it went through) based upon the information about the scattered photon.

- Mira: I disagree with your conclusion. If the photon had very large wavelength compared to the distance between the slits, it would not matter if an electron scatters off a photon because diffraction will limit our ability to resolve length scales smaller than the wavelength of the photon. In this case, scattering does not provide information about which slit the electron went through. For example, due to diffraction, one cannot use an optical microscope to examine viruses because their size is smaller than the shortest wavelength of visible light.

- Nancy: I agree with Mira that you may not be able to resolve two things by using photons of a wavelength larger than the length you are trying to resolve. In this context, if we are using photons with a wavelength larger than the distance between the slits, from the point of view of a photon, those two slits overlap and could be regarded as indistinguishable. If instead, the wavelength of the photon is smaller than the distance between the slits, a photon which scatters off an electron at one slit or another can provide information about which slit the interference occurred.

Do you agree with Pria and/or Mira and Nancy? Explain your reasoning.

Additionally, the tutorial has a checkpoint at the end for this question after answering this question and others related to how the wavelength of the photons affects the interference pattern, and below we include an excerpt from the summary for why the interference pattern remains unchanged when the photon wavelength is significantly larger than the distance between the slits.

- Each scattering event between an electron and a photon localizes the electron (for an instant) in a region of length scale comparable to the wavelength of the photon.
- But the wavelength of the photon is larger than the distance between the slits and, therefore, while the electron is at the slits, it is not localized in a region small enough to be able to tell with certainty that it is at one slit or at another.
- Therefore, for the case in which the wavelength of the photons emitted by the lamp is large compared to the distance between the slits, the interference pattern will remain essentially unchanged (compared to when the lamp is turned off).

The guided inquiry-based approach helps students learn why the interference pattern vanishes when the wavelength of the photons is significantly smaller than the distance between the slits in a similar manner, except that each scattering event between an electron and a photon localizes the photon in a region small enough such that it can be determined whether the electron is at one slit or another.

One of the ways in which the tutorial strives to help students learn to reason in terms of WPI is to help students make the connection between resolution (which is a prerequisite for understanding whether WPI is known in a given DSE setup) with the concept of WPI as it relates to the DSE. For example, the discussion above between Pria, Mira, and Nancy is an example of scaffolding to help students make connections between resolution and WPI. We note that it is possible that students had difficulties with resolution, and although the pretest and post test questions did not explicitly probe student understanding of resolution, in the post test we found that nearly all undergraduate students (who were motivated to learn from the QuILT via grade incentives) correctly interpreted the effect of photon wavelength on the interference pattern (91%, see Table VI) which suggests that the scaffolding provided helped students learn the concept of WPI, central to understanding various DSE setups (which include the addition of a monochromatic lamp).

*Learning objective 3.*—Recognize that the intensity of the lamp only affects the pattern observed on the screen when the wavelength of the photons emitted by the lamp is significantly smaller than the distance between the slits.

This learning objective is a corollary of learning objective 2. As discussed in the previous section, many students struggle to correctly incorporate the wavelength of photons emitted by the lamp. This difficulty results in the difficulty of correctly incorporating the intensity of the lamp from which photons are emitted and scatter off the particles in the DSE. In particular, many students expect that reducing the intensity of the lamp will result in changing the interference pattern regardless of whether the wavelength of the photons is significantly larger or smaller than the distance between the slits. After students learn about the photon wavelength dependence of whether interference is observed or not, the second part of the tutorial focuses on helping students utilize what they learned about the wavelength dependence to make predictions about the pattern that will form on the screen based upon the intensity of the lamp from which photons are emitted that scatter off the electrons. The students first consider the limiting cases of 100% intensity (meaning that every electron scatters off a photon) and 0% intensity (meaning that none of the electrons scatter off a photon). They are then guided to think about intermediate cases in which only some of the electrons scatter off a photon.

For example, the following question asks students to incorporate a lamp with an intensity such that half of the electrons scatter off the photons (but scattered electrons still arrive at the screen) in the DSE and make a prediction about the pattern that will form on the screen.

Q1. (i) Consider a case in which the lamp has intermediate intensity such that half of the electrons do not scatter off photons. Which one of the following statements is correct if the wavelength of the photons emitted by the lamp is significantly less than the distance between the slits?
(a) The interference pattern will go away.
(b) The interference pattern essentially remains unchanged.
(c) The interference pattern is still visible, however, it is harder to discern because of reduced contrast.
(d) The interference pattern becomes easier to discern because of increased contrast.
(ii) Explain your reasoning for your answer in (i).

Students are then prompted to use the simulation to check their prediction and reconcile differences between their prediction and observation if any. For example, after running the simulation, students are asked the following question:

Q2. What happened to the interference pattern as you lowered the intensity? Is this observation consistent with your answer to the preceding question? If it is not, reconcile the difference between your prediction and observation.

The QuILT then provides guidance and scaffolding support and strives to help students develop a good grasp of foundational concepts in quantum mechanics using the concrete context of the DSE. After working on the QuILT, students are expected to be able to qualitatively reason about how a single particle can exhibit the properties of both a wave and a particle, and be able to determine the de Broglie wavelength of a particle based on its mass or energy. They should be able to describe how scattering between a photon and a particle can provide WPI depending on the wavelength of the photon and whether a particle can be localized over a distance significantly smaller than the distance between the slits depending on the situation, and also describe how measurement of a particle's position at the screen collapses the wave function. Students are also expected to be able to explain the role of the photons emitted by the lamp and how scattering between these photons and the incoming particles can affect the presence of an interference pattern at the screen. Students should be able to reason about whether or not scattered photons give WPI about the particles after passing through the slits based on the wavelength of the photons, and be able to incorporate the intensity of the lamp into their predictions about what fraction of the particles incident on the slits will create interference fringes on the screen.

## VII. EVALUATION OF THE QUILT

Once it was determined that the QuILT was effective in meeting the learning objectives in individual administration, it was administered to students in two upper-level undergraduate quantum mechanics courses ($N = 46$) and graduate students who were simultaneously enrolled in the first semester of a graduate-level core quantum mechanics course and a course for training teaching assistants (TAs) (two separate years, $N = 45$). First, the students were administered a pretest. After the students worked on the pretest, they worked through the warm up and the main part of the QuILT in groups. They were given one week to work through the rest of the QuILT (including the homework component) and then submit it to the instructor as homework. They were then given a post test in class. Any students who did not work through the QuILT for any reason were omitted from the post test data.

We decided to use the QuILT in an upper-level undergraduate quantum mechanics course (and a graduate-level TA training course) because at the institution where the study was carried out a modern physics course may not discuss the DSE (which was the case for one faculty member who regularly teaches the course who was asked if he was interested in using the tutorial), or, if the DSE is discussed, it would only include the basic DSE setup with particles with mass. However, the main focus of the QuILT focuses is to help students understand more complex setups which include a monochromatic lamp placed between the slits and the screen. Also, we looked at multiple modern physics textbooks (e.g., Serway, Moses, and Moyer) and we have not found any discussions of DSE setups with single particles that include a monochromatic lamp placed between the slits and the screen. Therefore, the QuILT is best utilized in an upper-level quantum mechanics course (or courses at higher levels).

The upper-level undergraduate students who were enrolled in a quantum mechanics course received full credit for taking the pretest; the tutorial counted as a small portion of their homework grade for the course and their post tests were graded for correctness as a quiz. In addition, the upper-level undergraduates were aware that topics discussed in the tutorial could also appear on future exams since the tutorial was part of the course material. The graduate students were enrolled in a TA training course along with the graduate level core quantum mechanics course. In the TA training course, the graduate students learned about instructional strategies for teaching introductory physics courses. They were asked to work through the QuILT in one TA training class to learn about the effectiveness of the tutorial approach to teaching and learning. It was considered that the graduate students would recognize the value of the tutorial approach better if they discussed tutorials on topics which they are familiar with but do not fully understand (as opposed to discussing tutorials in introductory physics for which many graduate students are likely to be experts). If graduate students engage with these tutorials, they can learn the topics discussed and understand the value of utilizing these tools as supplements to instruction. They were given credit for completing the pretest, tutorial, and post test. However, their scores did not contribute to the final grade for the TA training course (which was a pass or fail course).

The students' performance on the pretests and post tests administered before and after they worked through the tutorial were used to assess the extent to which the learning objectives outlined in the previous section were achieved. The pretest and post test questions involve the following situations (the entire pretest and post test is provided in the Appendix):

Question 1 (Q1) presents a DSE setup with single electrons and asks students to describe a situation in which the introduction of a lamp between the slits and the screen close to the slits would destroy the interference pattern (although the electrons still arrive at the screen). A correct response mentions that the wavelength of the photons emitted by the lamp must be smaller than the separation between the two slits in order to localize the incoming electron sufficiently close to one of the two slits so that when the electron arrives at the screen we have WPI about the slit the electron went through.

Question 2 (Q2) presents a DSE using sodium (Na) atoms and asks students to calculate the number density at a point $x$ on the screen and to describe the pattern observed after a large number of atoms reaches the screen. In the situation presented, the wavelength of the photons emitted by the lamp is significantly *smaller* than the slit separation, while the intensity of the lamp is such that each Na atom scatters off a photon (but still arrives at the screen). The correct number density is $N/2|\Psi_1(x)|^2 + N/2|\Psi_2(x)|^2$ and the pattern on screen is no interference, which may be reasoned using WPI.

Question 3 (Q3) repeats the setup described in Q2, but now the wavelength of the photons emitted by the lamp is significantly *larger* than the slit separation. The correct number density is $N/2[|\Psi_1(x)|^2 + |\Psi_2(x)|^2 + 2|\Psi_1(x)||\Psi_2(x)|\cos\Delta\phi]$ and the pattern on the screen is an interference pattern since the photons' wavelength is not small enough to localize the Na atoms sufficiently to provide WPI (about which slit each particle went through) after the scattering takes place.

Question 4 (Q4) and question 5 (Q5) repeat Q2 and Q3, but now the intensity of the lamp has been decreased so that only half of the Na atoms scatter off the photons. The correct number densities are $N/2[|\Psi_1(x)|^2 + |\Psi_2(x)|^2] + N/2|\Psi_1(x)||\Psi_2(x)|\cos\Delta\phi$ and $N/2[|\Psi_1(x)|^2 + |\Psi_2(x)|^2 + 2|\Psi_1(x)||\Psi_2(x)|\cos\Delta\phi]$, and the patterns are partial interference (only Na atoms that do not scatter a photon show interference) and full interference (scattering does not localize Na atoms sufficiently to give WPI), respectively. The parameters for the photons that scatter off the Na atoms in the DSE situations for Q2–Q5 are summarized in Table I.

Students' responses for Q1–Q5 were categorized based on the most common types of responses in order to identify specific difficulties the students may have (detailed analysis of student responses is included in Sec. VIII). Between 20% and 30% of the students were independently categorized by a second rater for each question or question pair and an interrater agreement of greater than 90% was obtained in all cases.

### A. Concept-based rubric

Student performance on the pretests and post tests was evaluated using a concept-based rubric which often used "holistic" scoring designed to assess student understanding of relevant concepts across multiple questions (as discussed below) in order to determine whether students had developed a coherent knowledge structure of the relevant foundational issues in quantum mechanics and had met the learning objectives outlined in Sec. VI C. For example, learning objective 2 focuses on helping students learn that changing the wavelength of the photons may alter the interference pattern formed by the particles incident on the slits and why that would be the case under certain conditions. Students' responses to Q2 and Q3 were scored together in order to determine whether the students recognize and explain why (i) changing the wavelength of the photons that interact with the particles incident on the slits alters the interference pattern and (ii) a short wavelength photon (compared to the distance between the slits) localizes the particles (e.g., sodium atoms) close to one slit or the other and, therefore, provides WPI, whereas a long wavelength photon does not. Similarly, Q4 and Q5 were scored together using the same criteria used to score Q2 and Q3. Thus, the concept-based rubric was aligned with the learning objectives outlined in Sec. VI C. A summary of the rubric used to grade Q1, Q2–Q3, and Q4–Q5 is shown in Table II.

Between 20% and 30% of the data collected were independently rated by two different researchers using the rubric for all questions or question pairs, and the interrater reliability was excellent (greater than 90% agreement). As an example of how the rubric was applied,

TABLE I. Summary of relevant properties of photons from the lamp that interact with Na atoms in the DSE pretest and post test for Q2–Q5.

|  | Short wavelength | Long wavelength |
|---|---|---|
| Full intensity | Question 2 | Question 3 |
| Half intensity | Question 4 | Question 5 |

TABLE II. Summary of the rubric used to evaluate student responses to Q1, Q2–3, and Q4–5, with a total of two points possible for Q1, eight points possible question pair Q2 and Q3, and eight points possible for question pair Q4 and Q5.

| Q1 | Possible scores |
|---|---|
| 1. Mention that scattering a photon localizes the particle and may provide WPI and destroy the interference pattern. | 1, 0 |
| 2. Mention that the wavelength of the photons must be smaller than the distance between the slits ($\lambda < d$) in order to provide WPI. | 1, 0 |
| Total points possible | 2 |
| Q2–Q3 or Q4–Q5 | Possible scores |
| 1. Mention that the photon wavelength is an important consideration in determining the pattern that forms on the screen. | 1, 0 |
| 2. Correctly interpret the effect of wavelength on the interference pattern. (1 point possible for each question.) | 2, 1, 0 |
| 3. Find different number densities for the two questions (whether or not they are correct). | 1, 0 |
| 4. Number densities are correct. (1 point possible for each question.) | 2, 1, 0 |
| 5. Number densities are consistent with patterns. (1 point possible for each question.) | 2, 1, 0 |
| Total points possible | 8 |

TABLE III. Transcribed responses of Student A and Student B to Q2 and Q3.

Student A
Q2    $(N/2)(|\psi_1|^2 + |\psi_2|^2)$
       No interference, even distribution of photons.
Q3    $(N/2)(|\psi_1|^2 + |\psi_2|^2)$
       Still no interference pattern since photons give path info for each electron.

Student B
Q2    $(N/2)(|\psi_1|^2 + |\psi_2|^2)$
       There will be no interference pattern—the lamp photons give each atom which-path information when scattering.
Q3    $(N/2)[|\Psi_1(x)|^2 + |\Psi_2(x)|^2 + |\Psi_1(x)||\Psi_2(x)|\cos\Delta\phi]$.
       There will be an interference pattern. If $\lambda_{photon}$ > slit width, the two slits are indistinguishable (unresolvable) from each other to the photon, so the photon cannot give which-path information upon scattering.

TABLE IV. Scores assigned for solutions to Q2 and Q3 written by Student A and Student B (shown in Table III) using the rubric (see Table II), with commentary explaining the scores in italics.

| | A | B |
|---|---|---|
| 1. Mention that the photon wavelength is an important consideration in determining the pattern that forms on the screen. | 0 | 1 |
| *Student A: Made no mention of wavelength and described the same pattern for both situations.* | | |
| *Student B: Specifically mentioned wavelength.* | | |
| 2. Correctly interpret the effect of wavelength on the interference pattern. (1 pt. for each question.) | 1 | 2 |
| *Student A: Described the correct pattern for Q2 but not Q3.* | | |
| *Student B: Described both patterns correctly.* | | |
| 3. Find different number densities for the two questions. | 0 | 1 |
| *Student A: Did not find different number densities for Q2 and Q3.* | | |
| *Student B: Found two different number densities for Q2 and Q3.* | | |
| 4. Number densities are correct. (1 pt. for each question.) | 1 | 2 |
| *Student A: Wrote the correct number density for Q2 but not for Q3.* | | |
| *Student B: Wrote the correct number densities for Q2 and Q3.* | | |
| 5. Number densities are consistent with patterns. (1 pt. for each question.) | 2 | 2 |
| *Student A: Number densities were both consistent with the patterns described.* | | |
| *Student B: Number densities were both consistent with the patterns described.* | | |
| Total score | 4 | 8 |

Table III includes examples of responses (transcribed) for Q2 and Q3 written by two students (referred to as student A and student B), and Table IV shows how the rubric was applied to score the two students' responses for Q2 and Q3. Note that a students' answer to Q3 is considered correct because the student includes the cross term which depends on the path length difference $\Delta\phi$, even if this cross term is off by a factor of 2.

Average normalized gain [75] is commonly used to determine how much the students learned and takes into account their initial scores on the pretest. It is defined as

$$\langle g \rangle = \frac{\langle S_f \rangle - \langle S_i \rangle}{100\% - \langle S_i \rangle},$$

where $\langle S_f \rangle$ is the average percent score of the class on the post test and $\langle S_i \rangle$ is the average percent score of the class on the pretest [75]. We calculated the average normalized gains for both the upper-level undergraduate and graduate students using this equation.

## VIII. RESULTS

In order to determine the extent to which the QuILT was effective in helping students develop a coherent understanding of these concepts and addressing issues discussed in Sec. V related to learning objectives 1–3, we compared students' performances on the pretest and post test and measured the improvement. Below, we discuss the findings.

### A. Reasoning in terms of which-path information

Question 1 was an open-ended question and asked students to describe a situation in which the introduction of a lamp would destroy the electron interference pattern on the screen and why that would be the case. Many students struggled with this question on the pretest and provided a variety of responses. The student responses were categorized into six possible categories, as shown in Table V. A student response can fall in more than one category, which is why the percentages do not necessarily add up to 100%.

TABLE V. Categorization of student responses to Q1 as a percent of total responses for undergraduate (U) and graduate (G) students on the pretest and post test. Responses which received full credit are marked in bold, and responses which received at least partial credit are underlined.

| Q 1 | A | B | C | D | E | F |
|---|---|---|---|---|---|---|
| U pre | **9%** | 13% | 33% | 20% | 20% | 9% |
| U post | **91%** | 80% | 0% | 0% | 0% | 0% |
| G pre | **14%** | 32% | 36% | 5% | 14% | 5% |
| G post | **64%** | 22% | 9% | 4% | 16% | 2% |

The responses in Table V are categorized as follows:

(A) Mention $\lambda < d$: A correct response mentioned that the wavelength of the lamp's photons should be shorter than the separation between the slits. The students in this category had demonstrated that they understood the role of photon wavelength in determining whether interference is observed on the screen or not. Credit was also given to students who described how scattering via a photon localizes the particles and alters their momenta.

(B) Mention which-path information: At least half credit was given to any students who mentioned that if WPI is known from the scattered photons, then the interference pattern vanishes even if they did not explicitly describe the connection between WPI and the wavelength of the lamp's photons. Learning objective 1 of the DSE QuILT was that students learn to reason in terms of WPI in order to make predictions about the patterns that form on the screen. Any response that mentioned WPI (or used reasoning related to knowing which slit the particle went through to reach the screen) is counted in category B, even if the response was included in another category, which is why the rows of Table V do not necessarily add up to 100%. We note that it can be inferred that students who mention the correct condition ($\lambda < d$—category A) likely understand that if this condition holds, WPI is known for all particles and therefore the interference vanishes even if they do not explicitly say anything about WPI. However, we only counted student responses in category B if they explicitly mentioned that in a situation in which WPI is known, interference will not be observed. Therefore, the percentage of students in category A may be more or less than that in category B.

(C) Scattering: The most common response on the pretest described any type of physical scattering of the electrons due to collisions with the photons destroying the interference pattern without mentioning the constraints on photon wavelength. For example, one student stated

*"If scattering occurs enough between the lamp photons & the particles, they will completely convolute the interference pattern so it will no longer be visible. The screen will simply appear completely lit up."*

Another student stated

*"The interference pattern will be destroyed if the lamp has high enough intensity to scatter off the electrons."*

The question specifically mentions that the photons scatter off the electrons, so the responses in this category were mostly restating the information provided in the question without providing any additional details about the scattering process and how it would impact the interference on the screen when the particles arrive there. The responses of students in this category do not provide any evidence that students understand the mechanisms involved in destroying the interference pattern in this situation.

(D) Photon-electron interference: Several students (mostly undergraduates) described situations in which the wavelengths and phases of the photon and electron were aligned in such a way that the two would destructively interfere. For example, one student noted:

*"For destructive interference to occur the phase (scattering angle) between the photon and the electron must be such that maxima of the photon's wavelength correspond to minima of the electron's wavelength and vice versa."*

It is interesting that students are treating the incident particles and the photons from the lamp as "waves" that can interfere with each other and annihilate each other. Students with these types of responses are potentially invoking the principle of superposition as though the photon and electron are identical particles and the crest of one particle's wave will cancel the trough of the other particle's wave. This hypothesis is confirmed from interviews with students who invoked such a notion.

(E) Other responses: Many responses in this category were too simplistic and did not fall into other categories. These students often claimed that whenever a lamp is present, the interference pattern on the screen will vanish (without mentioning anything about the scattering of the particles off the photons from the lamp). For example, one student stated

*"There will be an interference pattern when the light bulb is off. When the light bulb is on, there will not be interference."*

(F) Incomplete or No Response: This category also includes those who wrote "I don't know." We note that all the students were given sufficient time to complete both the pretest and the post test and nearly all the students submitted their tests voluntarily. So if a student left a question blank, it is very likely that he or she did not know how to answer that question. Also, occurrences in which a particular question was left blank, but a subsequent question was answered were also fairly common, especially in the pretest, thus indicating that students most likely did not know how to answer the questions they left blank.

Table V shows that on the pretest 9% of undergraduates and 14% of graduate students were able to correctly identify the photon wavelength condition for whether an interference pattern will form on the screen. On the post test 91% of undergraduates and 64% of graduate students received full credit for their responses. As shown in Table V, 80% of undergraduate students explicitly used reasoning involving WPI to answer question 1 on the post test, compared to 13% on the pretest. These results demonstrate that the QuILT was effective in achieving learning objective 1 for a majority of students by addressing their initial difficulties with reasoning in terms of WPI. One possible explanation for the discrepancy between undergraduate and graduate students' post test scores in this regard is that the graduate students may be less motivated to engage with the QuILT due to the fact that (unlike the undergraduates) the graduate students were not graded for correctness on the post test and this material was not part of their other exams since there was no letter grade in the TA training course. Note that, however, these first year physics graduate students were also simultaneously enrolled in their first semester of a two semester core quantum mechanics course simultaneously although this material was not part of that course and that course was very traditional and did not focus on conceptual understanding of foundational concepts as in the QuILT.

Also, as shown in category D of Table V, on the pretest, about 20% of undergraduate students and 5% of graduate students described how the interference pattern on the screen will disappear if destructive interference occurs between the electrons and the photons from the lamp. None of the undergraduate students and only 4% of the graduate students used this reasoning on the post test.

### B. Difficulty recognizing the effect of photon wavelength on the interference pattern

Student responses to Q2 and Q3 were considered together, as were Q2 and Q4, and Q3 and Q5. The responses for these pairs were divided into the following six categories:

| (A) | Patterns and number densities are both correct. |
|---|---|
| (B) | Patterns are correct, but not the number densities. |
| (C) | Patterns are different and incorrect. |
| (D) | Patterns are the same and incorrect. |
| (E) | Other responses. |
| (F) | Incomplete or no response. |

Student responses to Q2 and Q3 were scored together to determine the extent to which learning objective 2 was achieved and students understood what will happen in the experiment if the wavelength of the photons emitted by the lamp is altered. For Q2, the wavelength of the photon is significantly smaller than the distance between the two slits (which localizes the particles incident on the slits sufficiently and impacts the interference pattern), while for Q3, the wavelength is significantly larger than the distance between the two slits (so the localization due to scattering does not give WPI for the particles incident on the slits in this case and interference is observed on the screen). The breakdown of the student responses to this question pair is shown in Table VI.

(A) Patterns and number densities correct.—Table VI shows that graduate students were more likely than undergraduates to respond correctly to question pair 2–3 on the pretest (25% vs 2%, respectively). On the post test, however, 91% of undergraduates answered correctly compared to only 71% of the graduate students.

TABLE VI. Categorization of undergraduate and graduate student responses to Q2 and Q3 as a percent of total responses.

| Q 2,3 | (A) | (B) | (C) | (D) | (E) | (F) |
|---|---|---|---|---|---|---|
| U pre | **2%** | 30% | 26% | 20% | 0% | 22% |
| U post | **91%** | 9% | 0% | 0% | 0% | 0% |
| G pre | **25%** | 5% | 30% | 20% | 5% | 16% |
| G post | **71%** | 2% | 9% | 13% | 4% | 0% |

(B) Only patterns correct.—Table VI shows that about 30% of undergraduate students on the pretest had a correct qualitative understanding of the role of photon wavelength in question pairs Q2–Q3 but did not know how to correctly represent the number densities in different situations (depending upon whether the interaction with the photons localized the particles sufficiently and there was WPI for the particles that arrived at the screen).

(C) Patterns different, incorrect.—Students in this category understood (or guessed) that changing the wavelength of the photons should change the pattern observed on the screen, but were not sure what that change should be.

(D) Patterns the same, incorrect.—Table VI shows that in the pretest, 20% of undergraduate and graduate students did not realize that changing the photon wavelength from significantly smaller to significantly larger than the distance between the slits will alter the pattern observed on the screen. Interestingly, 13% of graduate students on the post test maintained that the two patterns should be the same. They either did not think that changing the photon wavelength should affect the interference pattern, or did not make an effort to distinguish between the two situations.

(E) Other responses.—Some students, particularly graduate students, drew pictures that may or may not have represented interference patterns in the researchers' view, and a few of them wrote "Yes" or "No" for their responses without any elaboration. Since researchers did not understand what those responses meant even though there was an attempt to answer the questions, they were classified in this category.

(F) Incomplete or no response.—About 22% of undergraduates and 16% of graduate students did not fully respond on the pretest, or simply wrote "I don't know."

### C. Difficulty recognizing the effect of lamp intensity on the interference pattern

The pretest responses to question pair Q2 and Q4 and question pair Q3 and Q5 were assessed using the same categories as for Q2 and Q3 in the previous section to investigate learning objective 3, which is to understand the role of lamp intensity in determining the interference pattern on the screen. The categorization of responses to Q2 and Q4 is shown in Table VII. The students whose responses were placed in category (D) either failed to recognize that the intensity of the lamp would affect the pattern on the screen or did not make an effort to distinguish between the two situations in Q2 and Q4. About 26% of undergraduate and 18% of graduate students claimed that the patterns on the screen would be the same for both of these questions on the pretest.

TABLE VII. Categorization of undergraduate and graduate student responses to Q2 and Q4 as a percent of total responses.

| Q 2,4  | (A)     | (B)     | (C)  | (D)  | (E) | (F)  |
|--------|---------|---------|------|------|-----|------|
| U pre  | **9%**  | 20%     | 24%  | 26%  | 0%  | 22%  |
| U post | **88%** | 7%      | 5%   | 0%   | 0%  | 0%   |
| G pre  | **25%** | 7%      | 23%  | 18%  | 5%  | 23%  |
| G post | **56%** | <u>16%</u> | 9%   | 13%  | 4%  | 2%   |

As shown in category (D) of Table VII, about 13% of graduate students on the post test incorrectly maintained that the patterns should be the same in Q2 and Q4. None of the undergraduate responses manifest this mistake on the post test, even though about one-fourth of the undergraduate students had made this mistake on the pretest. This type of dichotomy in the performance of the undergraduate and graduate students demonstrates that the QuILT was more effective in helping undergraduate students learn to account for lamp intensity than the graduate students.

Question pair 3 and 5 present a situation in which the intensity of the lamp is altered while the wavelength of the photons is significantly larger than the distance between the slits such that scattering between the photons and atoms (the incident particles) will not affect the pattern on the screen. Student responses to these questions were compared and categorized, as shown in Table VIII. Correct responses are again in bold and partially correct are only underlined.

In Table VIII, responses in categories (A) and (B) indicate that many students understood or correctly guessed that the intensity of the lamp does not matter in this situation since the wavelength of the photons is not small enough to localize particles sufficiently to provide WPI. While about 94% of undergraduates recognized this fact on the post test, only about 51% of the graduate students did so.

As shown in category (C) of Table VIII, about one-third of undergraduates on the pretest did not realize that photons with wavelengths longer than the distance between the slits cannot alter the interference pattern, regardless of the intensity of the lamp. However, Table VIII shows that the percentage of undergraduates whose responses fell in category (C) was significantly lower in the post test. Interestingly, the percentage of graduate students who made this mistake and thought that the patterns should be different in Q3 and Q5 on the post test was actually slightly higher than the percentage on the pretest. The persistence of this difficulty with question pair Q3 and Q5 especially among graduate students on the post test illustrates a powerful phenomenological primitive, i.e., if you change something in the input, it should change *something* in the output [73]. However, in this case, changing the intensity of the lamp has no effect on the pattern. Prior research suggests that when students do not have a robust knowledge structure in a particular domain, it is common for students to use phenomenological primitives such as this [73] due to their prior conceptions. For example, Newton's third law of motion is a difficult concept for introductory students, and in the context of a small car and a large truck colliding head-on, many students claim that the truck exerts a larger force on the car than the car exerts on the truck. This is often due to the phenomenological primitive that "bigger means more," and since the truck has the larger mass it must therefore exert a larger force. The students' difficulty with the role of lamp intensity is specifically addressed in the QuILT to help them reason that while in some cases changing the intensity may impact the interference pattern, in other cases it has no impact. The fact that only 7% of undergraduate students made this error in the post test, but a comparable number of graduate students used this primitive on both the pretests and post tests suggests that many graduate students may not have engaged with the QuILT as effectively as the undergraduates.

We note that in Table VIII, in the case of Q3 and Q5, category D includes students who noted that the same pattern forms, but that pattern is incorrect. For example, they may incorrectly claim that when the wavelength of the photons emitted by the lamp is significantly larger than the distance between the slits, no interference pattern forms and answer that in both Q3 and Q5, no interference pattern forms. This answer is incorrect (the correct answer is that the pattern is the same in that an interference pattern forms in both cases). In contrast, in the other two tables (Tables VI and VII), any responses which have different interference patterns are incorrect. In order to use the same category (e.g., category D) in all tables, we have chosen that the responses included in category D are that the same interference pattern forms, but *the patterns are incorrect*. We do not need to explicitly specify that the patterns in category D are incorrect for Q2–Q3 or for Q2–Q4 in Tables VI and VII (because it is understood given that the correct answer is "different patterns form"). In particular, in Tables VI and VII, students who stated that the patterns are the same would be incorrect regardless of whether they noted that an interference pattern is observed or not. In particular, we specified this for Q3–Q5, because in that case, students may or may not be correct when they state that the pattern is the same in both situations (depending on their claim related to the presence or absence of an interference pattern). Similar comments apply to the other categories.

TABLE VIII. Categorization of undergraduate and graduate student responses to Q3 and Q5 as a percent of total responses.

| Q 3,5  | (A)     | (B) | (C) | (D) | (E) | (F) |
|--------|---------|-----|-----|-----|-----|-----|
| U pre  | **4%**  | 24% | 35% | 9%  | 0%  | 28% |
| U post | **80%** | 14% | 7%  | 0%  | 0%  | 0%  |
| G pre  | **14%** | 5%  | 34% | 18% | 5%  | 25% |
| G post | **49%** | 2%  | 40% | 4%  | 4%  | 0%  |

## IX. OVERALL STUDENT PERFORMANCE ON THE PRETEST AND POST TEST

The average scores on the pretests and post tests for the undergraduate and graduate students are shown in Fig. 5. We also calculate average normalized gains [75], $p$ values, and effect sizes in the form of Cohen's $d$ [$d = (\mu_1 - \mu_2)/\sigma_{pooled}$, where $\mu_1$ and $\mu_2$ are the averages of the two groups being compared and $\sigma_{pooled} = \sqrt{(\sigma_1^2 + \sigma_2^2)/2}$, where $\sigma_1$ and $\sigma_2$ are the standard deviations of the two groups], using individual group means and standard deviations. While the graduate students on average performed significantly better than the undergraduate students on the pretest (44% vs 23%, respectively, $p = 0.005$, $d = 0.43$), they performed significantly worse than the undergraduate students on the post test (73% vs 95%, respectively, $p < 0.001$, $d = 0.67$). Undergraduate students' average normalized gains were near the ceiling ($g = 0.94$), while the graduate students' corresponding gains were much lower ($g = 0.51$).

Figure 6 shows the distribution of the pretest and post test scores for each of the 45 undergraduate students (represented by blue triangles) and 46 graduate students (represented by red diamonds). The solid diagonal line through the middle of the plot represents the same score on the pretest and post test, so that all data points located above that line represent students who performed better on the post test than the pretest. The dotted lines located above and below the solid line represent the range of post test and pretest scores that were within 20% of each other. While nearly half of the graduate students had scores within this range (20 out of 45 students), only 3 of the undergraduate students had post test scores that were within 20% of their pretest scores. Moreover, those 3 undergraduate students already had pretest scores that were greater than 70% to begin with.

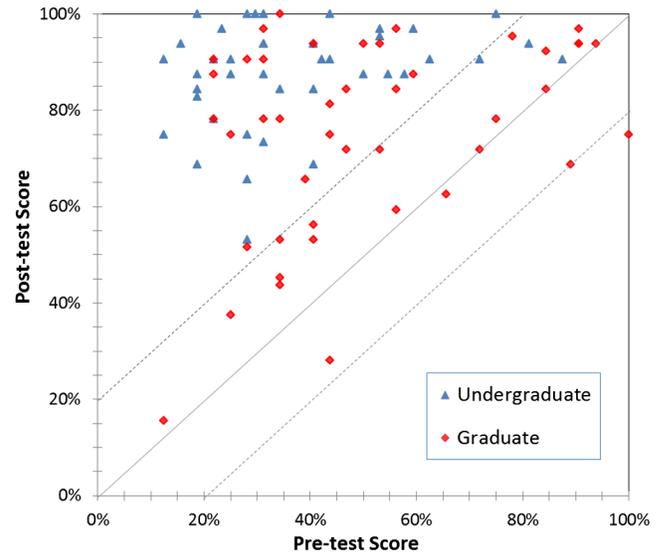

FIG. 6. Individual student post test scores vs pretest scores for undergraduate and graduate students. The solid diagonal line represents the cutoff for students whose post test scores were higher than their pretest scores. The dotted diagonal lines located above and below the solid diagonal line indicate cutoffs for students whose post test scores were within ±20% of the corresponding pretest score.

Figure 7 shows a histogram of the individual normalized gains for the undergraduate and graduate students, with dashed lines representing the average normalized gains for each group. Most undergraduate students had normalized gains greater than 0.7, and only two of them had normalized gains below 0.4. However, those two students scored very high on both the pretest and post test. Compared to the undergraduate students, the graduate students had more variation in their normalized gains. For example, 13 of the graduate students had normalized gains of 0.4 or less, compared to only two of the undergraduate students. (Note that four graduate students with negative normalized gains are not included in the histogram.)

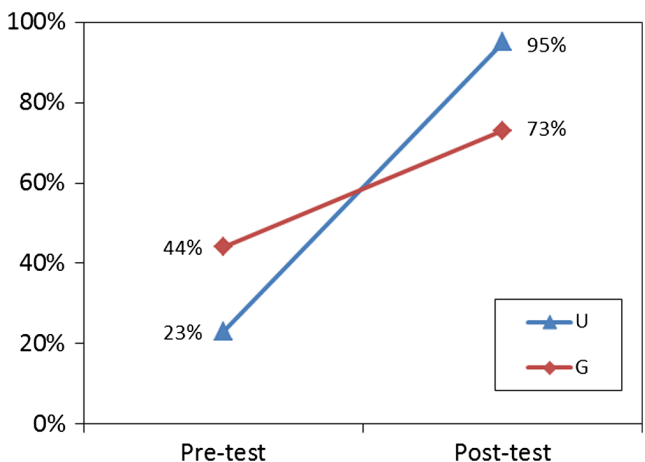

FIG. 5. Average pretest and post test scores for undergraduate (U) and graduate (G) students.

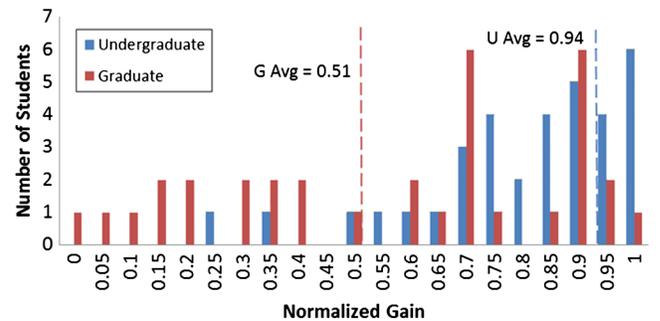

FIG. 7. Individual normalized gains for undergraduate students (blue bars) and graduate students (red bars), with average undergraduate and graduate student normalized gains represented with blue and red dashed lines, respectively.

TABLE IX. Average pretest and post test percentages on Q1, Q2–Q3, and Q4–Q5 for undergraduates (U) and graduate students (G), with $p$ values and effect size Cohen's $d$ for comparison of undergraduates and graduate students (the $p$ values and effect size are in the last two rows for each vertical comparison). Also, listed are the $p$ values and effect sizes for the difference between the means of the pretest and post test for each question or question pairs for each group, which show that while differences between the pretest and post test are significant in all cases, the effect sizes are significantly larger for undergraduates.

|   | Q 1 | | | | Q 2–3 | | | | Q 4–5 | | | |
|---|---|---|---|---|---|---|---|---|---|---|---|---|
|   | Pre | Post | $p$ | $d$ | Pre | Post | $p$ | $d$ | Pre | Post | $p$ | $d$ |
| U | 16 | 94 | <0.001 | 2.29 | 34 | 97 | <0.001 | 2.78 | 19 | 95 | <0.001 | 2.52 |
| G | 47 | 68 | 0.016 | 0.37 | 49 | 83 | <0.001 | 0.71 | 35 | 69 | <0.001 | 0.71 |
| $p$ | <0.001 | <0.001 | | | 0.018 | 0.005 | | | 0.023 | <0.001 | | |
| $d$ | 0.60 | 0.57 | | | 0.37 | 0.44 | | | 0.35 | 0.73 | | |

The average undergraduate and graduate student scores for questions Q1, Q2–Q3, and Q4–Q5 are shown in Table IX, with $p$ values and effect sizes for various comparisons. Note that Q2 and Q3 were graded together according to the rubric described in Sec. IV, as were Q4 and Q5. On average, graduate students performed better than the undergraduates in the pretest on all questions or question pairs and the reverse was true for the comparison of the post test scores for these two groups (as seen from the $p$ values and effect sizes $d$ in the last two rows in Table IX for each vertical comparison). A $t$-test comparison also indicated that the difference between the means of the pretest and post test for each question or question pairs is significant for each group (undergraduate and graduate students) but the effect sizes are significantly higher for the undergraduate students ($d$ is 2.29 for Q1, 2.78 for Q2–Q3, and 2.52 for Q4–Q5 for undergraduates). Note that in educational interventions large effects are considered to occur for Cohen's $d$ of 0.8 or more. The effect sizes for undergraduate students in this study are three times larger than that.

The QuILT was administered to both groups (undergraduates and graduate students) over a short time frame (pretest and post test for each group were separated by one week) without any additional in-class instructions on these topics. While there are other possible frameworks through which the differences between undergraduate and graduate student performances from pretest to post test may be interpreted, the impact of grade incentive is one of them. In particular, since other aspects of implementation were similar in both courses, one possible reason for the post test score discrepancy is that, as noted earlier, the undergraduates had grade incentives to learn from the QuILT while the graduate students worked on the QuILT in a TA training course with no final exam on which these types of questions could show up and a pass or fail grading scheme.

Some graduate students may have been less cognitively engaged in learning from the QuILT since it was graded only for completeness. We hypothesize that many students are not intrinsically motivated to learn even in advanced physics courses, and grade incentives for learning may provide the needed external motivation.

## X. SUMMARY

We investigated student difficulties with quantum mechanics concepts pertaining to the double-slit experiment in various situations that appear to be counterintuitive and contradict classical notions of particles and waves. We developed and carried out an evaluation of a research-validated QuILT that makes use of an interactive simulation to improve student understanding of the double-slit experiment and to help them develop a better grasp of foundational issues in quantum mechanics.

Data comparing the pretest and post test scores of upper-level undergraduate and graduate students indicate that the DSE QuILT was effective in improving students' understanding of these concepts that defy classical intuition such as wave-particle duality, effect of quantum measurement on the wave function, and explanation of whether interference should be observed after a large number of single particles pass through the slits. The QuILT strives to help students develop a coherent understanding of foundational concepts in various situations involving the DSE and helps students reason about whether or not interference of single particles is observed at the screen in the DSE in various situations. For example, when the photons from the lamp scatter off the particles at the slits, many students initially had difficulty understanding the effects of wavelength of the photons and intensity of the lamp on the interference pattern at the screen formed by single particles incident on the slits. About one-fifth of undergraduate students noted on the pretest that the photon wave and electron wave would somehow destructively interfere with each other during the scattering if their wavelengths were comparable, but none of the undergraduate students used this reasoning in their responses on the post test.

Moreover, upper-level undergraduates outperformed physics graduate students in the post test, although the reverse was true in the pretest. One possible reason for this difference may be the level of engagement with the QuILT due to the grade incentive. In the undergraduate course, the post test was graded for correctness, while in the graduate course it was graded for completeness.

## ACKNOWLEDGMENTS

We thank the National Science Foundation for Grant No. PHY-1505460, and Klaus Muthsam for the simulation that was adapted for the QuILT [48]. We are also thankful to various members of the department of physics and astronomy at the University of Pittsburgh (especially R. P. Devaty and E. Marshman) for helpful conversations and suggestions during the development of the tutorial.

## APPENDIX: DSE PRETEST

This is the full text of questions Q1 through Q5 on the DSE QuILT pretest and post test (which were identical). In the paragraph below "the figure below" refers to Fig. 1 in this article.

In questions 1–5, assume that particles are sent one at a time from the particle source. The figure below shows a double-slit experiment which was modified by adding a lamp (light bulb) between the double slit and the screen. The lamp is slightly off to the side so it does not block the slits. Assume that when the lamp is turned on, if scattering occurs between a particle used in the double-slit experiment and a photon from the lamp, this scattering occurs at the slits only.

• Assume that ALL the particles scattered by photons still reach the screen.

• Assume that a particle only scatters a single photon, i.e., multiple scattering is neglected.

1. Suppose you perform a double-slit experiment with electrons while the lamp is turned off and observe an interference pattern on the screen. You then repeat the experiment with the lamp turned on (assume that the intensity of the lamp is such that every particle used in the experiment scatters off a photon).

(i) Describe a situation in which this addition of the lamp between the double slit and the screen destroys the interference pattern observed on the screen (in the situation you describe, assume that all particles reach the screen even if scattering occurs between the particles and the photons emitted by the lamp).

(ii) Explain your reasoning for your answer in 1(i).

**Questions 2–5 refer to the following setup:**

You perform a double-slit experiment using Na atoms and observe an interference pattern on the screen. You then change the experiment by adding a lamp as discussed earlier.

• If slit 2 is closed, the wave function of a Na atom that goes through slit 1 and arrives at a point $x$ on the screen is $\Psi_1(x)$. If instead, slit 1 is closed, the wave function of a Na atom that goes through slit 2 and arrives at a point $x$ on the screen is $\Psi_2(x)$.

• For this example, if slit 2 is closed, and a total number $N$ of particles arrives at the screen, the number density of the particles at a point $x$ on the screen is $N|\Psi_1(x)|^2$.

• For questions 2–5, both slits are open.

2. For (i) and (ii) below, suppose that the wavelength of the photons is significantly smaller than the distance between the slits and the intensity of the lamp is such that each Na atom scatters off a photon. Also, assume that all the scattered atoms still reach the screen.

(i) Write down an expression for the number density of Na atoms at a point $x$ on the screen in terms of $\Psi_1(x)$ and $\Psi_2(x)$ after a large number $N$ of Na atoms arrive at the screen.

(ii) Describe the pattern you expect to observe on the screen after a large number $N$ of Na atoms have arrived at the screen. Explain your reasoning.

3. For (i) and (ii) below, suppose that the wavelength of the photons is significantly larger than the distance between the slits and the intensity of the lamp is such that each Na atom scatters off a photon. Also, assume that all scattered atoms still reach the screen.

(i) Write down an expression for the number density of Na atoms at a point $x$ on the screen in terms of $\Psi_1(x)$ and $\Psi_2(x)$ after a large number $N$ of Na atoms arrive at the screen.

(ii) Describe the pattern you expect to observe on the screen after a large number $N$ of Na atoms have arrived at the screen. How, if at all, is this pattern different from the pattern in 2(ii)? Explain your reasoning.

4. For (i) and (ii) below, suppose that the wavelength of the photons is significantly smaller than the distance between the slits and the intensity of the lamp is such that about half of the Na atoms scatter off a photon. Also, both slits are open and all the atoms reach the screen, including the ones that scatter.

(i) Write down an expression for the number density of Na atoms at a point $x$ on the screen in terms of $\Psi_1(x)$ and $\Psi_2(x)$ after a large number $N$ of Na atoms arrive at the screen.

(ii) Describe the pattern you expect to observe on the screen after a large number $N$ of Na atoms have arrived at the screen. How, if at all, is this pattern different from the pattern in 2(ii)? Explain your reasoning.

5. For (i) and (ii) below, suppose that the wavelength of the photons is significantly larger than the distance between the slits and the intensity of the lamp is such that about half of the Na atoms scatter off a photon. Also, both slits are open and all the atoms reach the screen, including the ones that scatter.

(i) Write down an expression for the number density of Na atoms at a point $x$ on the screen in terms of $\Psi_1(x)$ and $\Psi_2(x)$ after a large number $N$ of Na atoms arrive at the screen.

(ii) Describe the pattern you expect to observe on the screen after a large number $N$ of Na atoms have arrived at the screen. How, if at all, is this pattern different from the pattern in 3(ii)? Explain your reasoning.

---